\documentclass[letterpaper, 10 pt, conference]{ieeeconf} 
\IEEEoverridecommandlockouts                              
\usepackage{graphicx} 

\usepackage[hyphens]{url}
\usepackage{xurl}  
\usepackage[breaklinks=true]{hyperref}
\hypersetup{
	colorlinks=true,
	linkcolor=black,
	citecolor=black,
	filecolor=black,
	urlcolor=black,
}

\usepackage{cleveref}
\crefformat{footnote}{#2\footnotemark[#1]#3}

\usepackage{flushend}

\usepackage{microtype} 

\usepackage{xcolor}
\definecolor{bg}{rgb}{0.95,0.95,0.97}
\usepackage[outputdir=build, frozencache,cachedir=.]{minted}
\newenvironment{juliacode}{%
\VerbatimEnvironment
\begin{minted}[breaklines, autogobble, bgcolor=bg, fontsize=\footnotesize]{julia2.py:Julia2Lexer -x}%
}%
{%
\end{minted}%
}%

\newenvironment{ccode}{%
\VerbatimEnvironment
\begin{minted}[breaklines, autogobble, bgcolor=bg, fontsize=\footnotesize]{c}%
}%
{%
\end{minted}%
}%

\newmintinline{julia}{}
\AtBeginEnvironment{snugshade*}{\vspace{-\FrameSep}} 
\AfterEndEnvironment{snugshade*}{\vspace{-\FrameSep}}

\newcommand{\footnotelink}[1]{\footnote{\href{https://#1}{#1}}}
\newcommand{\package}[1]{\texttt{#1}}
\newcommand{\packagelink}[2]{\href{#2}{\texttt{#1}}\footnote{\label{#1}\href{https://#2}{#2}}}

\title{C codegen considered unnecessary: go directly to binary, do not pass C.\\
Compilation of Julia code for deployment in model-based engineering.}
\author{Fredrik Bagge Carlson, Cody Tapscott, Gabriel Baraldi, Chris Rackauckas \\ \emph{JuliaHub Inc.}}
\date{November 2024}

\begin{document}

\maketitle

\begin{abstract}

    Since time immemorial an old adage has always seemed to ring true: you cannot use a high-level productive programming language like Python or R for real-time control and embedded-systems programming, you must rewrite your program in C. We present a counterexample to this mantra by demonstrating how recent compiler developments in the Julia programming language allow users of Julia and the equation-based modeling language \package{ModelingToolkit} to compile and deploy binaries for real-time model-based estimation and control. Contrary to the approach taken by a majority of modeling and simulation tools, we do not generate C code, and instead demonstrate how we may use the native Julia code-generation pipeline through LLVM to compile architecture-specific binaries from high-level code. This approach avoids many of the restrictions typically placed on high-level languages to enable C-code generation. As case studies, we include a nonlinear state estimator derived from an equation-based model which is compiled into a program that performs state estimation for deployment onto a Raspberry Pi, as well as a PID controller library implemented in Julia and compiled into a shared library callable from a C program.
\end{abstract}

\section{Introduction}

Model-based engineering (MBE) has emerged as a cornerstone of modern engineering, enabling the efficient design, analysis, and deployment of complex systems. By representing systems as mathematical or graphical models, engineers can simulate, test, and validate designs prior to physical implementation, reducing development time and cost. Central to the MBE workflow is the use of high-level modeling languages to describe systems, followed by the automatic generation of low-level C code for deployment on embedded hardware or other real-time systems.

Although this workflow has proven highly effective, it presents several issues. Code generation capabilities often come with a significant price tag, presenting a financial incentive to find competitive alternatives. Furthermore, high-level modeling languages with code-generation capabilities often impose restrictions on expressiveness to ensure compatibility with C-code generation, such as limitations on which functions may be called,\footnotelink{se.mathworks.com/help/simulink/slref/coder.screener.html} use of variable-sized arrays,\footnotelink{se.mathworks.com/help/coder/ug/limitations-with-variable-size-support-
for-code-generation.html}\footnotelink{specification.modelica.org/master/arrays.html} and use of flexible data structures, operator overloading, and arrays of objects.\footnotelink{se.mathworks.com/help/simulink/ug/how-working-with-matlab-classes-
is-different-for-code-generation.html} Additionally, integrating preexisting libraries or leveraging cutting-edge algorithms and hardware accelerators in these environments can be cumbersome, as it requires significant manual effort to make them compatible with code generation, or the hardware support is missing etc. These limitations can hinder the pace of development and stifle innovation, particularly in domains that require flexibility or powerful computing.

A fully-featured high-level programming language that seamlessly compiles to fast, efficient machine code offers a compelling alternative in situations when C-code generation is not strictly required.\footnote{C-code generation may be \emph{required} by regulation, safety standards or by constraints imposed by the target platform.} Such a language would preserve the benefits of high-level expressiveness while enabling rapid prototyping, integration with advanced libraries, and direct deployment on real-time systems. Julia \cite{Bezanson2017}, a free and open-source programming language increasingly popular for technical computing, is well-positioned to fulfill this role. Known for its combination of high-level syntax and performance comparable to low-level languages, Julia allows developers to write expressive, memory safe, and maintainable code without sacrificing execution speed.

Historically, however, Julia lacked support for compiling small, standalone binaries or shared libraries, a feature critical for deploying code to resource-constrained environments such as embedded systems. Recent advancements in Julia’s compiler technology have started addressing this gap, enabling a new approach to MBE workflows. In this paper, we demonstrate how these developments, combined with the \package{ModelingToolkit} equation-based (acausal) modeling framework, enable the compilation and deployment of Julia programs directly to the target system without relying on C-code generation. By leveraging Julia's native code generation pipeline, we eliminate the restrictions typically imposed by traditional workflows, offering greater flexibility and enabling the use of Julia's rich ecosystem of already existing libraries. To illustrate this capability, we design a simple equation-based model and a nonlinear state estimator, compile the resulting program, and deploy it on a Raspberry Pi for real-time state estimation. We also convert a Julia package implementing PID controllers into a C-callable shared library.

\section{Compilation of Julia code}

Typical languages used for embedded programming perform ahead-of-time (AOT) compilation by either strongly restricting the code a user may write (via their static type system, à la C/C++ or Java) or by emitting code that supports runtime-dynamic types, which tends to result in much lower performance. Julia typically circumvents this trade-off via its JIT (just in time) compiler, which allows it to emit machine code with sufficient specialization for high performance without statically restricting user code. While utilizing a JIT leads to good performance without restricting dynamicity, JITs aren't suitable for most embedded applications because of their large binary sizes and the undesirable latency characteristics of generating code on-the-fly.

To support ahead-of-time compilation without a JIT compiler, Julia includes a dedicated "trimming" mode, described in the following section.

\subsection{Trimming Support}
The trimming mode enforces that all code emitted is \emph{sufficiently specialized} for high performance. In this context, sufficient specialization refers to the compiler being able to infer a bounded number of potential targets for each function call. The compiler additionally prunes any loaded code not reachable from an entry point as well as any code associated with the development environment, such as for the REPL (read-eval-print-loop).

Trimming is performed through a static analysis on a list of user-declared \texttt{entrypoints}, which will become the exported symbols for the compiled library/executable. The static analysis flags locations where code is too dynamic to achieve sufficient performance, and if no such problems are found the compiler emits a small, high-performance binary. Currently, this typically results in a lower limit on the binary size of a few megabytes for common MBE applications (results for example applications are presented in \cref{sec:case_studies}). The produced binaries currently need to link to the Julia runtime library which currently adds 48MB to the total size.\footnote{A large part of this number is due to the inclusion of openBLAS.} The result is still a dramatic size reduction of several hundred megabytes compared to the amount of code loaded for traditional interactive Julia development, and the binary size produced by the pre-existing deployment solution \package{PackageCompiler.jl}. Trimming is not yet performed on the Julia runtime itself, and we foresee future work bringing the size of this down significantly. 

%

\subsection{Emulation for cross-compilation} \label{sec:cross-compilation}

Another significant challenge for code generation on embedded targets relates to handling CPU architecture (arch) and Operating System (OS) differences. To support these use cases, the Julia compilation workflow can be executed via an emulator such as \texttt{qemu}, allowing an x86-64 macOS or Windows machine to generate libraries and executables for, e.g., an AArch64 Linux target such as the Raspberry Pi 4.

The usage of an emulator allows for powerful compile-time code-execution capabilities, similar to \texttt{constexpr} in a language like C++, including constructing compile-time hashmaps and other data structures or pre-computing mathematical functions. Additional work is also ongoing to support cross compilation natively via the Julia compiler, eliminating the need for an emulator.

\section{Case studies} \label{sec:case_studies}
\subsection{Case study 1: Model-based state estimation}

To demonstrate the practical utility of compiling Julia programs for model-based estimation, we present a case study that implements a state estimator for a continuously stirred tank reactor (CSTR) using the \packagelink{ModelingToolkit.jl}{docs.sciml.ai/ModelingToolkit/dev/} (MTK) framework. This case study showcases the entire workflow, from model development to deployment as a standalone binary, and highlights the flexibility and efficiency enabled by Julia's recent compiler advancements. The following steps outline the process:

\subsubsection{Modeling the System}
We begin by constructing a simple acausal model of a CSTR using MTK. While the modeling is not the primary subject of this tutorial and is left out for brevity, we encourage the interested reader to scan through the accompanying software repository \cite{repo}. The model captures the dynamic behavior of the reactor, represented in the form $\dot{x} = f(x, u, p, t)$, where $x$ is the state vector, $u$ represents control inputs, $p$ denotes parameters, and $t$ is time. MTK provides a high-level interface for describing these dynamics, which are then symbolically transformed into an efficient numerical representation.

\subsubsection{Extracting the Dynamics Function}
Once the model is defined, we generate a Julia function that computes the system dynamics. The function adheres to the $\dot{x} = f(x, u, p, t)$ structure (customization options exist), enabling its use in various downstream tasks, including state estimation, model-based control loops and deployment in the form of a Functional Mock-up Unit (FMU).

Being a symbolic language, MTK performs code generation from the symbolic representation to \emph{Julia code} (C-code generation for this step is also available). In a normal simulation workflow, MTK will immediately hand the generated Julia code to the Julia compiler without exposing it to the user, but in the present use case we will instead redirect the generated Julia code to a file for inclusion in the compiled binary. This approach has several benefits; it allows the model translation and subsequent Julia code generation to be separated from the compilation of the downstream use of the model into machine code, saving time in repeated applications of the final compilation step. It also allows the model translation and Julia code generation, which are steps that require installing and loading somewhat heavy libraries, to be performed on any machine, not necessarily the target hardware.

\subsubsection{Designing the State Estimator}
For state estimation, we employ an Unscented Kalman Filter (UKF), leveraging the preexisting package \packagelink{LowLevelParticleFilters.jl.}{baggepinnen.github.io/LowLevelParticleFilters.jl/dev/} The continuous-time system dynamics generated by MTK are discretized using a fixed-step Runge-Kutta 4 (RK4) integrator for use in the UKF.

To make the state estimator performant and amenable to compilation using \package{juliac}, we ensure that all arrays used by the dynamics calculations and the UKF are statically sized.\footnote{This is a common procedure in a language like C which may be optionally performed in Julia as well.} We use the compiler extension packages \packagelink{JET.jl}{github.com/aviatesk/JET.jl} and \packagelink{AllocCheck.jl}{https://github.com/JuliaLang/AllocCheck.jl} to verify these properties and provide rich feedback about potential problems during the development of the application code.

We place the code that configures and instantiates the UKF in top-level scope in a Julia module, and let the created UKF object be a global \juliainline{const} within this module. The main entry point of the program reads input and measurement data from files and performs the filtering using the preexisting function \juliainline{forward_trajectory} from \package{LowLevelParticleFilters.jl}:
\begin{juliacode}
Base.@ccallable function main()::Cint
    y = reinterpret(SVector{4, Float64}, read("data_y.bin")) 
    u = reinterpret(SVector{2, Float64}, read("data_u.bin"))
    length(y) == length(u) || error("Data-length mismatch")
    println(Core.stdout, "Data length ", length(y))

    # Perform filtering
    sol = forward_trajectory(kf, u, y)
    println(Core.stdout, "loglik = ", sol.ll)
    return 0
end
\end{juliacode}

The full code is available in the file \package{juliac\_demo.jl} in the repository \cite{repo}.

\subsubsection{Compiling and Deploying the Binary}
To deploy the state estimator, we use the \package{juliac} compiler tool to compile the Julia code into a standalone binary. The resulting program includes a \texttt{main} function that reads measurement data from files, executes the state estimation algorithm along the data trajectory, and computes the log-likelihood of the observed data given the model.

The binary is reasonably lightweight and can run on resource-constrained hardware, such as a Raspberry Pi (RPi) (tested with RPi 4 model B 4GB), making it suitable for real-time applications. In order to produce a binary for the RPi platform, we perform the compilation directly on the RPi. See \cref{sec:cross-compilation} for details on how to perform cross compilation instead.

The binary produced was 3.5MB and running it was 2.484s / 26.667ms $\approx$ 93x faster than running the Julia script that defines the UKF and performs the filtering using the standard Julia execution model.\footnote{Benchmark performed on a desktop PC, the run time of the compiled binary on the Raspberry Pi is about 4x slower.}
\begin{juliacode}
    using BenchmarkTools
    # Benchmark standard execution model
    @btime run(`julia --project juliac_demo.jl`)

    # Benchmark compiled executable
    @btime run(`./juliac_demo`)

    stat("juliac_demo").size / 1024^2 # 3.495
\end{juliacode}

The standard Julia execution model takes significantly longer to run the application due to performing the compilation just ahead of running it, while the binary compiled using \package{juliac} is completely ahead-of-time compiled.

\subsection{Case study 2: A PID controller shared-object library}
This example is aimed at demonstrating how instead of an executable binary, we may compile a shared library from a Julia package. The example will produce a shared library from \package{DiscretePIDs.jl} \cite{DiscretePIDs-repo}, a Julia package implementing a standard PID controller, with functions for updating parameters while ensuring \emph{bumpless transfer} and other features commonly added to practical PID implementations.

To ensure that the \package{DiscretePIDs.jl} Julia package is suitable for real-time applications, it is verified to not perform any GC-managed heap allocations or runtime dispatch during operation. This verification is carried out using the compiler tool \package{AllocCheck.jl} and is performed as part of the continuous integration suite of \package{DiscretePIDs.jl}.

All user-interfacing functions in \package{DiscretePIDs} are written taking an instance of a PID-controller object as an argument. To produce a compiled library with a simple interface, we create a constant global instance of a controller and introduce wrapper functions with the same name as each library function, in which the global controller instance is implicitly passed along. The simplified API exposed to the user this way allows the user to change the PID parameters, so this simplification does not limit the flexibility of the compiled library in any other way than restricting the use to a single controller instance only. Each wrapper function is marked as \juliainline{@ccallable} to indicate that the function should be included in the shared library.

The code snippet below illustrates how the global controller instance and one of the wrapper functions are created.
\begin{juliacode}
const T = Float64 # The numeric type used

# Global const controller instance
const pid = DiscretePID(;
    K  = T(1),      # proportional gain
    Ti = 1,         # Integral time
    Td = false,     # Derivative time
    Ts = 1,         # Sample time
)

# Wrapper function with implicit pid object
@ccallable function calculate_control!(r::T, y::T, uff::T)::T
    DiscretePIDs.calculate_control!(pid, r, y, uff)
end
\end{juliacode}

The shared-library file produced by \package{juliac} was 1.7MB large, not including the Julia runtime library which any loading code must link to separately. To demonstrate use of the shared library from a C program, we implement a simple program (abridged for brevity):
\begin{ccode}
int main() {
    void *lib_handle = dlopen(LIB_PATH, RTLD_LAZY);
    // verify success ...

    jl_init_with_image_t jl_init_with_image = (jl_init_with_image_t)dlsym(lib_handle, "jl_init_with_image");
    calc_ctrl_t calc_ctrl = (calc_ctrl_t) dlsym(lib_handle, "calculate_control!");
    set_K_t set_K = (set_K_t) dlsym(lib_handle, "set_K!");
    // load more julia functions ...
    // Verify success ...

    // Init julia
    jl_init_with_image(JULIA_PATH, LIB_PATH);

    // Trivial program computing ctrl outputs and modifying K
    double r = 1.0, y = 0.0, uff = 0.0;
    double result = calc_ctrl(r, y, uff);
    printf("calc_ctrl! returned: %f\n", result);
    result = calc_ctrl(r, y, uff);
    printf("calc_ctrl! returned: %f\n", result);
    set_K(0.0, r, y);
    for (int i = 0; i < 3; ++i) {
        result = calc_ctrl(r, y, uff);
        printf("calc_ctrl! returned: %f\n", result);
    }

    return 0;
}
\end{ccode}

The entire example is available in the repository \cite{DiscretePIDs-repo}.

\section{Current limitations and future work}

While the discussion above demonstrates the ability for high-level Julia-defined applications to be compiled for embedded real-time applications, there are currently a few remaining threads that need to be addressed in order for this approach to be a complete solution for embedded applications. First of all, this approach currently uses an unreleased version of Julia based on the prerelease of Julia v1.12, and thus requires a non-standard installation. Secondly, the workflows from above do not showcase how to perform cross-compilation, which currently requires an emulator for the target device in which the compilation must occur. The Julia ecosystem at the time of writing provides no special tools for this cross-compilation and this step may currently be challenging to perform.

Additionally, it should be noted that not all hardware can be targeted via the current approach. Emitted code is generally high-performance, but it relies on the presence of a traditional Operating System (e.g. Linux, Windows, etc.). It also requires the Julia runtime, which means that it is limited to supporting the same architecture and OS combinations that Julia typically supports. ARMv8 and x86-64 are well-supported, but other targets like ARMv7 and RISC-V have limited support or are currently experimental. In particular, bare metal and RTOS targets are not supported today. These environments typically come with strong restrictions on what a program can do, including prohibiting dynamic memory allocation or lacking a traditional filesystem, which the Julia runtime doesn't yet support.

\section{Conclusion}
We have shared the potential of using recent compiler developments in the Julia ecosystem for accelerating MBE workflows involving code deployment in situations where C-code generation is not strictly necessary. Our examples demonstrate production of both executable binaries and C-callable shared libraries from Julia code, enabling engineers to leverage the combination of the equation-based modeling language \package{ModelingToolkit} and preexisting Julia libraries in their workflow without rewriting anything in C. The \package{juliac} compiler tool is still in development, and we foresee that rapid improvements during the coming year will make this a powerful tool for accelerated MBE going forward.

\bibliographystyle{IEEEtran}
\bibliography{references}

\begin{thebibliography}{1}
\providecommand{\url}[1]{#1}
\csname url@rmstyle\endcsname
\providecommand{\newblock}{\relax}
\providecommand{\bibinfo}[2]{#2}
\providecommand\BIBentrySTDinterwordspacing{\spaceskip=0pt\relax}
\providecommand\BIBentryALTinterwordstretchfactor{4}
\providecommand\BIBentryALTinterwordspacing{\spaceskip=\fontdimen2\font plus
\BIBentryALTinterwordstretchfactor\fontdimen3\font minus
  \fontdimen4\font\relax}
\providecommand\BIBforeignlanguage[2]{{%
\expandafter\ifx\csname l@#1\endcsname\relax
\typeout{** WARNING: IEEEtran.bst: No hyphenation pattern has been}%
\typeout{** loaded for the language `#1'. Using the pattern for}%
\typeout{** the default language instead.}%
\else
\language=\csname l@#1\endcsname
\fi
#2}}

\bibitem{Bezanson2017}
J.~Bezanson, A.~Edelman, S.~Karpinski, and V.~B. Shah, ``Julia: A fresh
  approach to numerical computing,'' \emph{SIAM review}, vol.~59, no.~1, pp.
  65--98, 2017.

\bibitem{repo}
F.~Bagge~Carlson, ``static\_kalman,''
  \url{https://github.com/baggepinnen/static_kalman}.

\bibitem{DiscretePIDs-repo}
------, ``{DiscretePIDs.jl},''
  \url{https://github.com/JuliaControl/DiscretePIDs.jl}.

\end{thebibliography}

\end{document}